 \documentclass[final,5p,times,twocolumn,authoryear]{elsarticle}

\usepackage{amsmath, amsfonts, amssymb} 

\usepackage{algorithm, algorithmic} 
\usepackage{listings} 

\usepackage{graphicx} 
\usepackage{xcolor} 
\usepackage{tikz} 
\usetikzlibrary{shapes.geometric, arrows} 

\usepackage{booktabs} 
\usepackage{longtable} 
\usepackage{makecell} 
\usepackage{tabularx} 
\usepackage{array} 
\usepackage{adjustbox} 
\usepackage{colortbl} 
\usepackage{multirow} 

\usepackage{textcomp} 
\usepackage{subfigure} 
\usepackage{bm} 
\usepackage{fancyvrb} 
\usepackage{multicol} 
\usepackage{enumitem} 

\usepackage{geometry} 
\usepackage{balance} 

\usepackage{hyperref} 
\usepackage[switch]{lineno} 
\usepackage[utf8]{inputenc} 
\usepackage{lipsum} 

\journal{Knowledge Based Systems}

\begin{document}

\begin{frontmatter}



\title{From Text to Returns: Using Large Language Models for Mutual Fund Portfolio Optimization and Risk-Adjusted Allocation}


\author[first]{Abrar Hossain\fnref{eq1}}
\ead{abrar.hossain@rockets.utoledo.edu}

\author[first]{Mufakir Qamar Ansari\fnref{eq2}}
\ead{Mufakir.Ansari@rockets.utoledo.edu}

\author[third]{Haziq Jeelani}
\ead{haziq.jeelani@cgu.edu}

\author[fourth]{Monia Digra}
\ead{monia@bennett.edu.in}

\author[first]{Fayeq Jeelani Syed\corref{cor1}}
\ead{sjeelan@rockets.utoledo.edu}

\fntext[eq1]{These authors contributed equally to this work.}
\fntext[eq2]{These authors contributed equally to this work.}

\cortext[cor1]{Corresponding author.}

\affiliation[first]{organization={Department of Electrical Engineering and Computer Science, University of Toledo},
            addressline={2801 W Bancroft St},
            city={Toledo},
            postcode={43606},
            state={OH},
            country={USA}}

\affiliation[third]{organization={Claremont Graduate University},
            addressline={150 E. 10th Street},
            city={Claremont},
            postcode={91711},
            state={CA},
            country={USA}}

\affiliation[fourth]{organization={School of Computer Science Engineering and Technology, Bennett University},
            addressline={Plot No. 8-11, Tech Zone II},
            city={Greater Noida},
            postcode={201310},
            state={Uttar Pradesh},
            country={India}}

\begin{abstract}
Generative AI (GenAI) is transforming diverse sectors, from healthcare to natural language processing, by solving complex, data-intensive problems with unprecedented efficiency. In finance, GenAI offers an immense potential in addressing critical challenges such as portfolio optimization and risk management—key pillars of modern finance that ensure strategic asset allocation and risk mitigation. These practices enable investors to balance returns while safeguarding investments against uncertainties, a task made increasingly complex by volatile market dynamics and growing datasets. This study situates itself at the intersection of GenAI and finance, leveraging the capabilities of large language models (LLMs) to redefine traditional financial decision-making.

This research examines the performance of advanced LLMs, including Microsoft Phi 2, Mistral 7B, and Zypher 7B, in generating actionable and risk-adjusted sectoral allocation strategies for mutual funds. A structured framework is employed, integrating retrieval-augmented generation pipelines with advanced optimization techniques to enhance the decision-making process. The methodology incorporates macroeconomic indicators to generate context-aware strategies. A comparative evaluation is done to present strengths and limitations of each model, focusing on predictive accuracy, risk management capabilities, and practical applicability. Scalability and ethical considerations are addressed throughout the study, with a particular emphasis on interpretability and compliance with financial regulations to ensure responsible deployment.

Among the evaluated models, Zypher 7B consistently outperforms its counterparts in maximizing returns and delivering superior risk-adjusted outcomes. Its ability to integrate complex, non-linear dependencies and contextual indicators positions it as a robust tool for financial optimization. The findings demonstrate a substantial improvement over baseline allocations, underscoring the transformative potential of generative AI in asset management. Furthermore, the study identifies critical areas for future exploration, including real-time adaptability, enhanced model transparency, and the integration of multi-objective strategies. By bridging the gap between generative AI and practical financial applications, this work provides a foundation for innovative, efficient, and adaptable solutions in asset management, offering actionable insights for both academia and industry stakeholders.
\end{abstract}




\end{frontmatter}




\section{Introduction}
\label{introduction}

Quantitative investing has become a cornerstone of modern financial management, utilizing quantitative signals derived from a wide range of data sources to design and optimize investment portfolios \citep{quant_investing}. These sources include traditional financial data, such as market prices, economic indicators, and corporate financial statements, as well as more complex, unstructured text-based information. Advances in natural language processing (NLP) have made it feasible to extract insights from text data, enabling more nuanced and adaptive quantitative investment strategies \citep{nlp_finance}. At the forefront of these advancements are Large Language Models (LLMs), which excel in interpreting and generating natural language, unlocking opportunities for integrating linguistic insights into financial decision-making \citep{llm_nlp_finance}.

Prominent LLMs, such as Microsoft Phi 2 \citep{microsoft2023phi2}, Mistral 7B \citep{jiang2023mistral}, Zypher 7B \citep{huggingface2023zephyr}, GPT-4 \citep{openai2023gpt4}, and PaLM 2 \citep{anil2023palm2}, have demonstrated exceptional capabilities in processing complex textual data and generating actionable insights. These models can be fine-tuned to address domain-specific tasks, such as portfolio optimization, where the goal is to maximize returns while minimizing risks and volatility. By integrating LLMs into financial workflows, investors can dynamically adjust sector-level allocations based on evolving market conditions and nuanced textual signals, such as earnings reports and macroeconomic forecasts. This integration enhances the use of unstructured data in data-driven investment strategies, paving the way for adaptive and efficient decision-making.

This study explores the application of open-source LLMs in optimizing mutual fund portfolios, focusing on their ability to recommend sector-level allocations that effectively balance risk and reward. Using a retrieval-augmented generation pipeline (Figure~1), financial data is transformed into vector embeddings for efficient context retrieval. These embeddings are then processed by LLMs to generate initial allocation recommendations, which are refined using blended optimization techniques incorporating metrics like the Sharpe ratio \citep{sharpe1966mutual}. The study evaluates the performance of Microsoft Phi 2, Mistral 7B, and Zypher 7B, analyzing their strengths in financial reasoning, contextual understanding, and decision-making. Additionally, it highlights how LLMs bridge the gap between unstructured textual insights and actionable financial strategies, expanding the scope of quantitative investing.

\subsection*{Contributions}
The main contributions of this paper are as follows:
\begin{enumerate}
    \item Propose a novel framework for mutual fund portfolio optimization using retrieval-augmented generation pipelines to leverage open-source LLMs for sector-level allocations.
    \item Evaluate and compare the performance of state-of-the-art LLMs (Microsoft Phi 2, Mistral 7B, and Zypher 7B) in financial optimization tasks, offering insights into their contextual reasoning capabilities and practical applications.
    \item Introduce a blended optimization approach integrating LLM-generated recommendations with financial metrics like the Sharpe ratio to achieve improved risk-adjusted returns.
    \item Demonstrate the feasibility of deploying LLMs in resource-constrained environments by adapting Microsoft Phi 2 for CPU-based operations.
\end{enumerate}

\section{Related Work}

\subsection{Machine Learning in Financial Portfolio Optimization}

The adoption of machine learning (ML) techniques in financial portfolio optimization has transformed asset management by addressing limitations of traditional methods. Classical approaches, such as the mean-variance framework introduced by Markowitz, often fail to capture the complexities and dynamic nature of financial markets, leading to suboptimal portfolios due to estimation errors. ML techniques, by contrast, leverage data-driven methods to enhance portfolio performance.

Notable advancements include regularization techniques, such as performance-based regularization (PBR), which constrain sample variances to improve portfolio stability \citep{ban2018machine}. Conditional Portfolio Optimization (CPO) represents another innovation, tailoring portfolio allocations to prevailing market conditions by conditioning optimization on various market features \citep{chan2023conditional}. Deep learning models, such as Long Short-Term Memory (LSTM) networks, have also been employed to predict asset returns, showing significant promise in constructing portfolios that outperform benchmarks during volatile market conditions \citep{fischer2018deep, martinez2024portfolio}.

Robo-advisors have further advanced ML applications, enabling scalable optimization algorithms capable of managing extensive asset universes. These developments provide efficient solutions to address the complexities of modern financial markets \citep{deng2023unified}. Additionally, ML techniques support automated rebalancing, predictive analytics, and dynamic risk management, facilitating more efficient and customized asset allocation \citep{cfa2024machine}.

\subsection{GeneAI in Financial Portfolio Optimization}

GenAI has opened new possibilities for harnessing textual data in portfolio optimization. Early research focused on word-level embeddings, which lacked the ability to model contextual relationships. Subsequent studies incorporated sentiment analysis from financial news and social media to predict stock prices, while attention mechanisms modeled text-driven numerical impacts on market movements.

The advent of LLMs has further revolutionized the use of text data in finance. Encoder-only models like BERT \citep{devlin2019bert} and DeBERTa \citep{he2021deberta} excel in learning contextual embeddings, while decoder-only models like GPT-3 \citep{brown2020gpt3} and Mistral specialize in sequential text generation. Encoder-decoder models, such as T5 \citep{raffel2020t5} and BART \citep{lewis2020bart}, combine these approaches for sequence-to-sequence tasks.

LLMs, pre-trained on vast text corpora, can be adapted to downstream tasks using prompt-based techniques or fine-tuning. Prompt-based approaches guide pre-trained models through crafted inputs, while fine-tuning adjusts their parameters for domain-specific tasks. Studies leveraging LLMs for financial sentiment analysis and predictive factor extraction have demonstrated their potential to bridge the gap between textual insights and quantitative investing.

\section{Limitations of Current Machine Learning Approaches}

Despite their advancements, traditional ML approaches in portfolio optimization face several challenges. (The reader might want to read more about this subject matter here: \citep{dixon2020financial, gu2020empirical, nguyen2019risk, dixon2021machine, fischer2018deep, tsantekidis2020financial}):

\begin{enumerate}
    \item \textit{Data Integration and Pattern Recognition}: ML models often struggle to integrate diverse datasets and uncover complex patterns, particularly with unstructured data \citep{kpmg2023genai}.
    \item \textit{Adaptability to Market Changes}: Static ML models are less responsive to evolving market conditions, leading to suboptimal decisions. GenAI facilitates scenario simulations for robust portfolio optimization \citep{leewayhertz2023genai}.
    \item \textit{Handling Unstructured Data}: Unstructured sources like financial news and social media pose challenges for ML models, while GenAI excels in extracting actionable insights \citep{leewayhertz2023genai}.
    \item \textit{Computational Efficiency}: Traditional ML methods can be resource-intensive, whereas GenAI offers efficient processing for large datasets \citep{kpmg2023genai}.
\end{enumerate}

\section{Problem Formulation}

Portfolio optimization involves determining the optimal allocation of weights \( \mathbf{w} = [w_1, w_2, \dots, w_n] \) across \( n \) sectors to maximize expected returns while minimizing associated risks, subject to practical constraints. Mathematically, this problem can be expressed as:

\[
\begin{aligned}
    & \max_{\mathbf{w}} \mathbb{E}[R(\mathbf{w})], \\
    & \text{s.t. } \sum_{i=1}^n w_i = 1, \quad w_i \geq 0, \quad \sigma(\mathbf{w}) \leq \tau.
\end{aligned}
\]

Here, \( \mathbb{E}[R(\mathbf{w})] \) represents the expected portfolio return as a function of the weight vector \( \mathbf{w} \), \( \sigma(\mathbf{w}) \) denotes the portfolio risk, defined as the standard deviation of returns, and
\( \tau \) is a predefined risk tolerance threshold.

Portfolio optimization relies on structured financial data, including:
\begin{itemize}
    \item \textit{Fund Attributes}: Fund identifiers, dates, average net asset values (NAVs), average percentage returns, and risk levels.
    \item \textit{Sectoral Exposures}: Allocation to sectors such as technology, healthcare, finance, and energy.
    \item \textit{Macroeconomic Indicators}: Average interest rates, inflation rates, and other market-level signals influencing asset behavior.
\end{itemize}

Traditional optimization methods often face challenges in handling the complexities of these datasets, particularly when balancing the trade-off between maximizing returns and minimizing risks under fluctuating macroeconomic conditions.

Generative AI, particularly Large Language Models (LLMs), offers a novel approach to addressing these challenges. By learning intricate relationships within structured financial data, LLMs can generate allocation recommendations \( \mathbf{w}^* \) that optimize risk-adjusted returns. This is achieved using the Sharpe ratio, which quantifies the trade-off between returns and risk:

\[
\mathbf{w}^* = \arg\max_{\mathbf{w}} \text{Sharpe Ratio} = \frac{\mathbb{E}[R(\mathbf{w})]}{\sigma(\mathbf{w})}.
\]

The application of LLMs in portfolio optimization presents several key challenges:
\begin{enumerate}
    \item \textit{Sectoral Weight Optimization}: Determining the optimal sectoral allocations \( \mathbf{w}_\text{sector} = [w_\text{tech}, w_\text{health}, w_\text{finance}, w_\text{energy}] \) that balance sector-specific risks and returns.
    \item \textit{Dynamic Macroeconomic Integration}: Incorporating macroeconomic indicators, such as interest rates and inflation, to enable real-time adjustments in portfolio allocations.
    \item \textit{Evaluation Metrics}: Assessing LLM effectiveness across multiple dimensions, including:
    \begin{itemize}
        \item \textit{Absolute Returns}: Direct financial gains from portfolio strategies.
        \item \textit{Risk-Adjusted Returns}: Performance as measured by metrics like the Sharpe ratio.
        \item \textit{Computational Efficiency}: The resource requirements of deploying LLMs for large-scale portfolio analysis.
    \end{itemize}
\end{enumerate}

This study aims to design and validate an LLM-driven framework for portfolio optimization. The framework:
\begin{itemize}
    \item Leverages structured financial data to generate robust, risk-aware sectoral allocation recommendations.
    \item Addresses the limitations of traditional optimization methods by incorporating nuanced relationships within structured and unstructured data.
    \item Enhances decision-making in mutual fund management through a dynamic, data-driven approach that adapts to evolving market conditions.
\end{itemize}

\section{Methodology}
\label{Methodology}

\begin{figure*}[htbp]
    \centering
    \label{fig:pipeline}\includegraphics[width=0.8\textwidth]{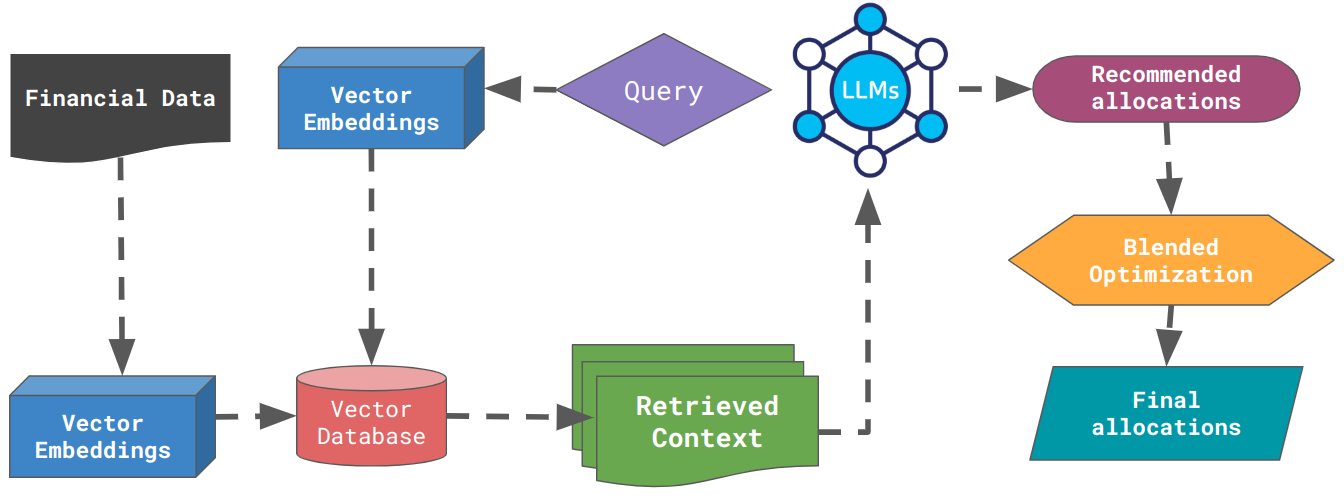} 
    \caption{Overview of the LLM-based blended optimization framework. The process begins with financial data being transformed into vector embeddings, which are stored in a vector database. When a query is made, relevant context is retrieved from the database and fed into Large Language Models (LLMs). The LLMs generate recommended sectoral allocations, which are further refined using a blended optimization approach to produce final allocations. This workflow integrates data-driven context retrieval, generative AI capabilities, and optimization techniques to improve portfolio performance.}
    \label{fig:two_column_example}
\end{figure*}

This study evaluates the performance of three open-source Large Language Models (LLMs)—Microsoft Phi-2, Mistral 7B, and Zypher 7B—in optimizing mutual fund portfolios. We take three funds for this study, namely: Fund A, Fund B, and Fund C. The models were tasked with providing sectoral allocation recommendations to maximize returns while minimizing risk and volatility. A structured approach was adopted to design, implement, and evaluate the optimization process for the selected models.

\subsection{Dataset Preparation}
A synthetic dataset was developed to simulate realistic mutual fund data, ensuring consistency and relevance for the optimization task. Each data entry included attributes such as:
\begin{itemize}
    \item \texttt{Fund\_ID},
    \item \texttt{Date},
    \item \texttt{Average\_NAV},
    \item \texttt{Average\_Return\%},
    \item \texttt{Risk\_Level}.
\end{itemize}
Additionally, sectoral exposures were incorporated, including:
\begin{itemize}
    \item \texttt{Technology\_Exposure\%},
    \item \texttt{Healthcare\_Exposure\%},
    \item \texttt{Finance\_Exposure\%},
    \item \texttt{Energy\_Exposure\%}.
\end{itemize}
Macroeconomic indicators such as \texttt{Average\_Interest\_Rate\%} and \texttt{Average\_Inflation\_Rate\%} were also included. The dataset was processed into structured documents using LangChain’s \texttt{Document} class, enabling compatibility with vector stores and query pipelines for interaction with LLMs.

\subsection{Model Selection and Configuration}
Three LLMs were selected based on their architectures and capabilities (see Table~\ref{tab:model_comparison}). Each model was evaluated for its ability to handle mutual fund optimization tasks under distinct computational and architectural constraints:

\begin{table*}[ht]
\centering
\caption{Details of Mistral 7B, Zypher 7B Beta, and Microsoft Phi-2. The table compares key attributes of the three Large Language Models (LLMs) used in this study. Mistral 7B, developed by Mistral AI, and Zypher 7B Beta, fine-tuned by Hugging Face, both feature 7 billion parameters and share an open-source licensing model (Apache 2.0 and MIT, respectively). Microsoft Phi-2, a lighter model with 2.7 billion parameters, is also Transformer-based and licensed under MIT. The comparison highlights architectural similarities while underscoring differences in model size and fine-tuning strategies.}
\label{tab:model_comparison}
\begin{tabular}{lccc}
\toprule
\textbf{Attribute} & \textbf{Mistral 7B} & \textbf{Zypher 7B Beta} & \textbf{Microsoft Phi-2} \\ 
\midrule
\textbf{Developer}   & Mistral AI         & Hugging Face            & Microsoft               \\ 
\textbf{Model Size}  & 7B                 & 7B                      & 2.7B                   \\ 
\textbf{Architecture} & Transformer-based  & Fine-tuned Mistral 7B   & Transformer-based       \\ 
\textbf{Licensing}   & Apache 2.0         & MIT                     & MIT                     \\ 
\bottomrule
\end{tabular}
\end{table*}

\begin{itemize}
    \item \textit{Microsoft Phi-2}: Released in June 2023, this transformer-based model includes 2.7 billion parameters and is optimized for CPU-efficient operations. Using the configuration \texttt{low\_cpu\_mem\_usage=True}, it was integrated into a \texttt{transformers} text-generation pipeline for resource-constrained environments. Microsoft Phi-2 balances computational efficiency with performance, making it ideal for scenarios with limited hardware resources \citep{phi2_reference}.

    \item \textit{Mistral 7B}: Developed by Mistral AI and released in September 2023, this model consists of 7 billion parameters and is optimized for tasks requiring advanced contextual understanding and reasoning. Custom prompts and parameter tuning were employed to align the model’s performance with mutual fund optimization requirements. Its Apache 2.0 license provides flexibility for academic and commercial applications \citep{mistral_reference}.

    \item \textit{Zypher 7B Beta}: Introduced by Hugging Face in January 2024, Zypher 7B Beta is a fine-tuned version of Mistral 7B with enhancements for financial decision-making tasks. While it shares the same model size (7 billion parameters), task-specific fine-tuning improves its accuracy and computational efficiency. Zypher 7B is distributed under an MIT license, offering broad flexibility for integration \citep{zypher_reference}.
\end{itemize}

Each model utilized a HuggingFace tokenizer to preprocess and encode text, ensuring efficient interaction with the data and the optimization pipeline. Their performance in sectoral allocation optimization was evaluated based on key metrics, including computational efficiency, risk-adjusted returns, and sectoral allocation accuracy.

\subsection{Prompt Design and Optimization}
A well-structured prompt template was designed to guide the LLMs in generating sectoral allocation recommendations. The template focused on providing detailed fund information in the \texttt{Context} field and requesting precise percentage adjustments for each sector based on risk levels and exposure. The prompt was carefully calibrated to elicit actionable and contextually relevant optimization strategies (see Listing~\ref{lst:prompt_template}).

\begin{lstlisting}[caption={Prompt Template for Sectoral Allocation Recommendations}, label={lst:prompt_template}, basicstyle=\ttfamily\footnotesize, breaklines=true, frame=single]
You are a financial assistant specializing in mutual fund portfolio optimization. Your role is to analyze user-provided data, apply advanced financial strategies, and recommend portfolio adjustments to maximize returns while considering the user's risk tolerance and investment goals.

Key behaviors and constraints:
1. Compliance: Always adhere to financial regulations and disclaimers, explicitly stating that you are not a licensed financial advisor and that users should consult professionals before making decisions.
2. User-Focused: Prioritize the user's preferences, such as risk tolerance, time horizon, and specific goals (e.g., growth, income, or capital preservation).
3. Clarity and Simplicity: Provide explanations that are clear and concise, including step-by-step rationale for your recommendations.
4. Actionable Advice: Offer specific adjustments to the portfolio, such as rebalancing allocations, diversifying across asset classes, or switching to funds with lower expense ratios.
5. Data Awareness: Provide recommendations based on the data you have been provided.

Always act in the user's best interest and ensure your recommendations are tailored to their input and objectives.
\end{lstlisting}

\subsection{Embedding and Vector Store}
To enable efficient retrieval of relevant fund data, embeddings were generated using the lightweight model \texttt{all-MiniLM-L6-v2}. The embeddings were stored in a Chroma vector store, configured to retrieve the top three most relevant documents per query. This configuration balanced computational efficiency with contextual depth, ensuring the LLMs received the most pertinent data for their analyses.

\subsection{Query and Response Pipeline} The updated query and response pipeline leverages financial data, vector embeddings, and large language models (LLMs) to optimize mutual fund allocations. The pipeline includes the following components:

\begin{itemize} 
    \item \textit{Financial Data Processing}: Raw financial data is processed into vector embeddings that encapsulate relevant features for optimization tasks. 
    \item \textit{Vector Database Storage and Retrieval}: The vector embeddings are stored in a vector database, enabling efficient retrieval of contextually relevant information based on user queries. 
    \item \textit{Query Handling and Context Retrieval}: User queries are transformed into vector embeddings, which are used to retrieve the most relevant context from the vector database. 
    \item \textit{LLM-Based Analysis and Recommendations}: The retrieved context is input into LLMs along with the query. The LLMs analyze the data and generate recommendations for sectoral or fund-specific allocations. 
    \item \textit{Blended Optimization}: The recommendations undergo a blended optimization process to refine allocations based on multi-objective constraints or user preferences. 
    \item \textit{Final Output}: The final optimized allocations are presented to the user as actionable insights for mutual fund management. 
\end{itemize}

This enhanced pipeline ensures efficient handling of large datasets, seamless integration of user queries, and high-quality allocation recommendations.

\begin{figure*}[t!]
\centering
\subfigure[]{\label{fig:kripke_two_param_heatmap}\includegraphics[width=0.45\textwidth]{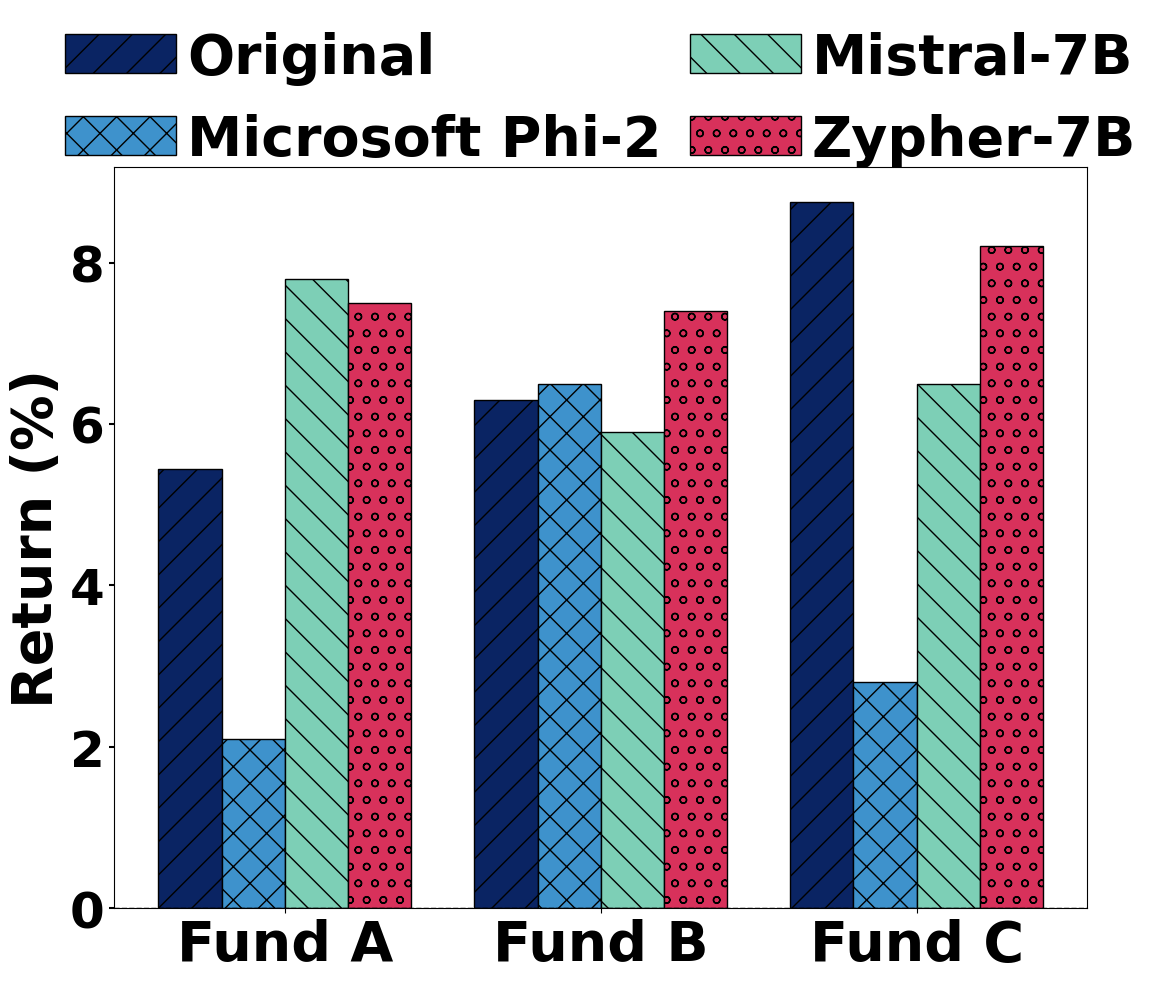}}\hspace{0.2cm}
\subfigure[]{\label{fig:varying_input_params_oracle}\includegraphics[width=0.45\textwidth]{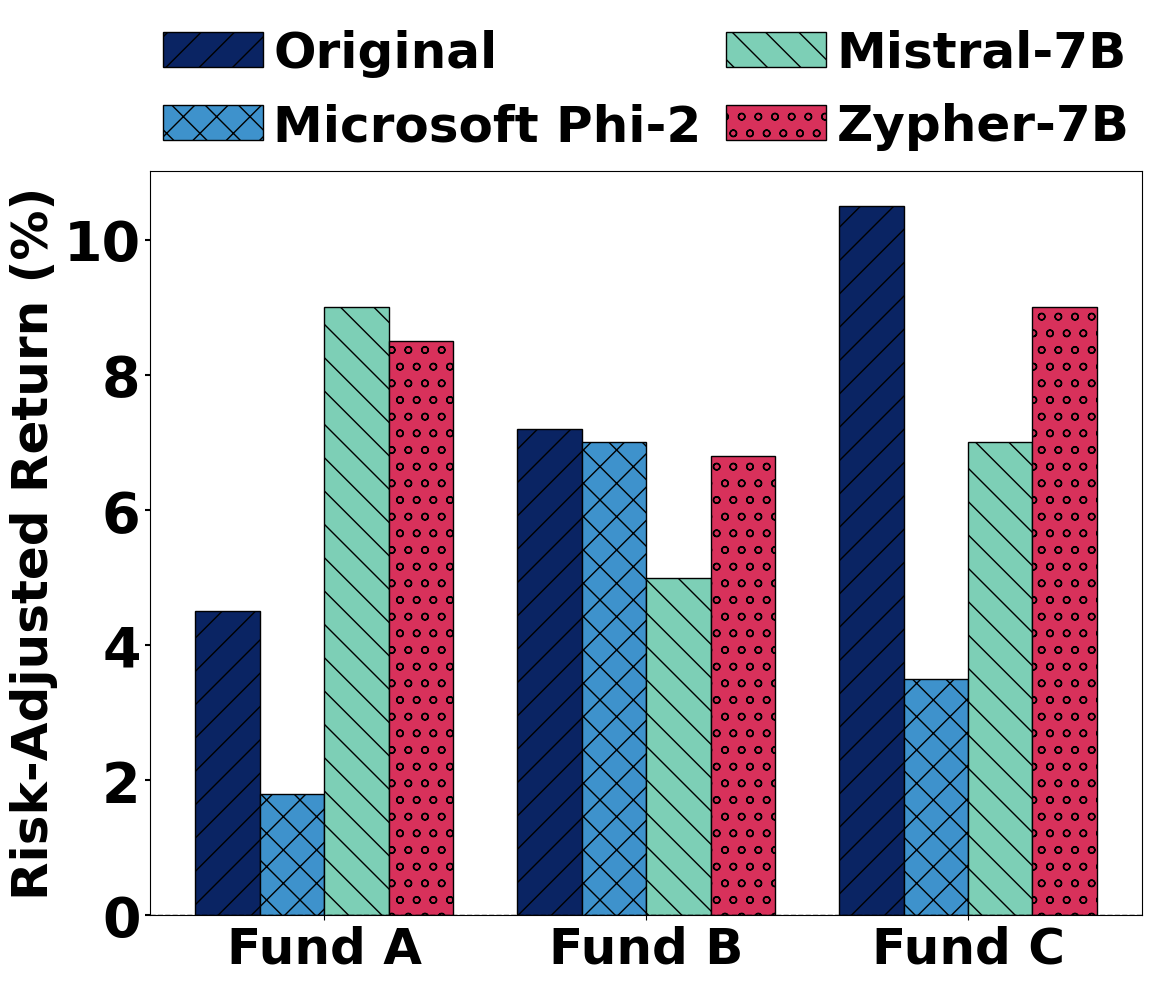}}
\vspace{-3mm}
\caption{Performance comparison of mutual fund portfolios optimized by Large Language Models (LLMs) — Microsoft Phi-2, Mistral-7B, and Zypher-7B — against the original allocations.
(a) Return (\%): Zypher-7B demonstrates the highest improvements in Fund A and Fund C, outperforming other models, while Mistral-7B performs well in Fund A. Microsoft Phi-2 underperforms across all funds.
(b) Risk-Adjusted Return (\%): Zypher-7B achieves superior risk-adjusted returns for Funds A and C, reflecting its strength in balancing returns and risks. Mistral-7B also shows competitive performance, whereas Microsoft Phi-2 struggles across all funds.}
\label{fig:kripke_execution1}
\end{figure*}

\begin{figure*}[t!]
\centering
\subfigure[]{\label{fig:kripke_two_param_heatmap}\includegraphics[width=0.45\textwidth]{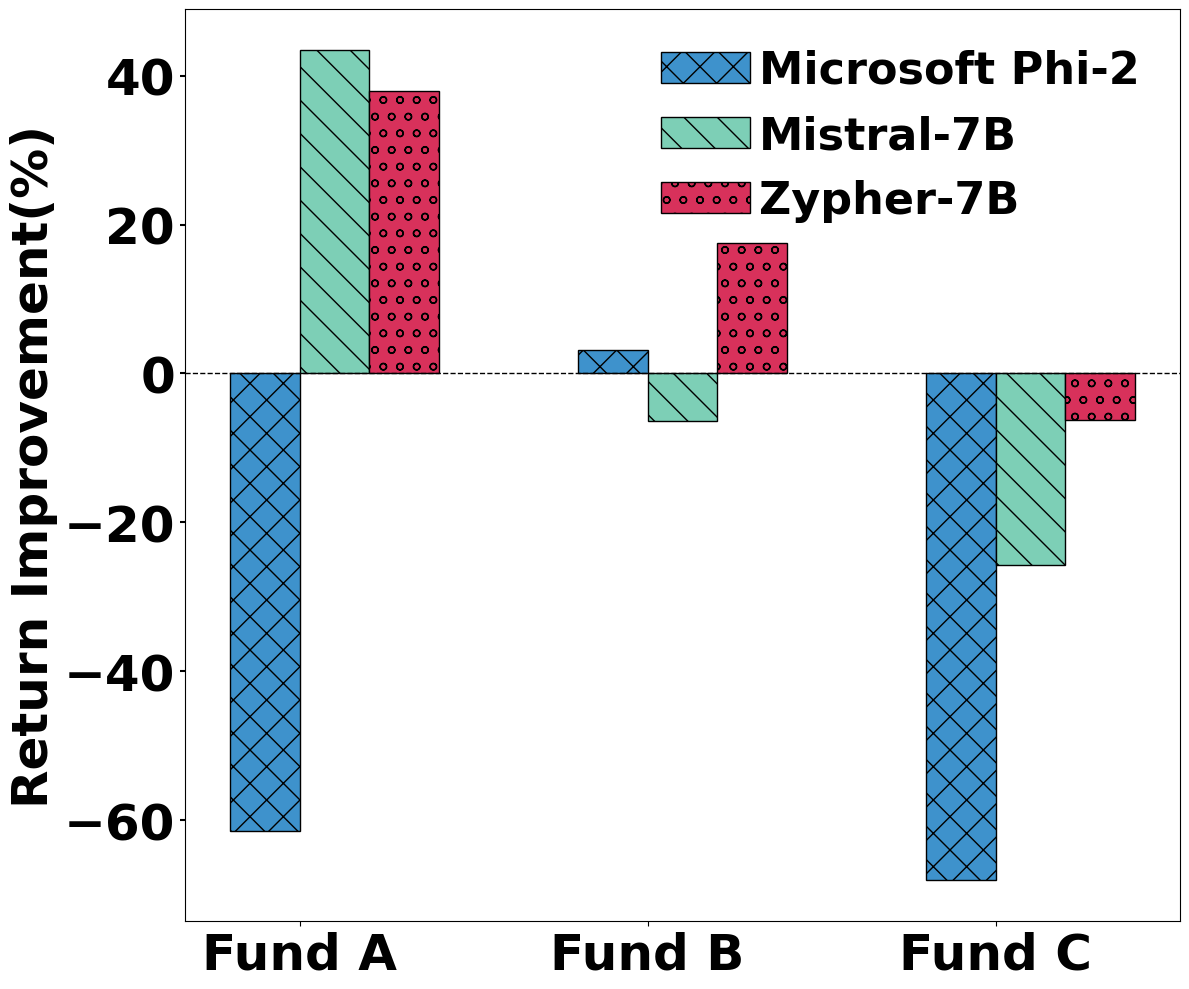}}\hspace{0.2cm}
\subfigure[]{\label{fig:varying_input_params_oracle}\includegraphics[width=0.45\textwidth]{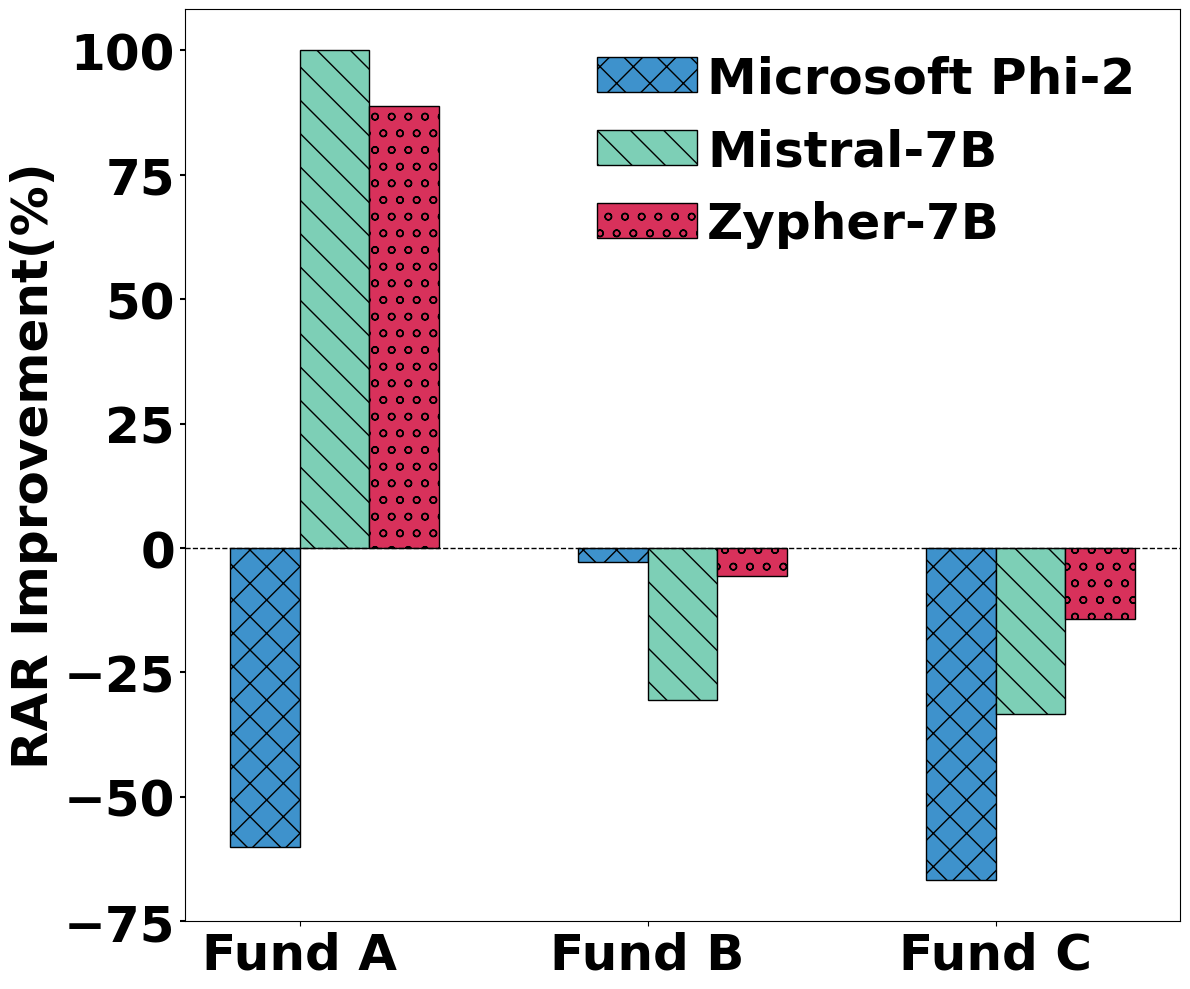}}
\vspace{-3mm}
\caption{(a) Return Improvement (\%): Comparison of return improvements across three models — Microsoft Phi-2, Mistral-7B, and Zypher-7B. Mistral-7B and Zypher-7B demonstrate significant positive improvements for Fund A, while Microsoft Phi-2 underperforms across all funds, particularly in Fund A and Fund C, where returns drop drastically.
(b) Risk-Adjusted Return (RAR) Improvement (\%): Mistral-7B and Zypher-7B achieve substantial RAR improvements in Fund A, with Mistral-7B reaching close to 100\%. However, Microsoft Phi-2 shows negative RAR improvements across all funds, indicating poor performance in balancing returns and risk.}
\label{fig:kripke_execution2}
\end{figure*}

\section{Evaluation and Discussion}
\label{Evaluation}

This study evaluates the performance of three open-source Large Language Models (LLMs)—Microsoft Phi 2, Mistral 7B, and Zypher 7B—in optimizing sectoral allocations for three mutual funds: Fund A, Fund B, and Fund C. Each model was tasked with reallocating sectoral exposures to maximize portfolio returns while minimizing volatility and overall risk. The optimized allocations were compared to the original allocations using key metrics, including absolute returns, risk-adjusted returns (e.g., Sharpe ratio), and volatility. The analysis provided insights into the models’ capabilities to balance risk and return trade-offs, offering a comprehensive evaluation of their suitability for financial optimization. The findings are discussed in detail below.

\subsection{Comparison Across Models}

The performance of Microsoft Phi 2, Mistral 7B, and Zypher 7B was evaluated using four key metrics: \textit{Return}, \textit{Risk-Adjusted Return}, \textit{Improvement on Return}, and \textit{Improvement on Risk-Adjusted Return}. Figures \ref{fig:kripke_execution1} and \ref{fig:kripke_execution2} present the comparative results, which are detailed as follows:

\paragraph{Return}
The models exhibited varying levels of success in enhancing portfolio returns. Microsoft Phi 2 consistently underperformed relative to the other models, particularly for Fund A, where the return was significantly lower. In contrast, Mistral 7B and Zypher 7B demonstrated superior performance, with Zypher 7B achieving the highest return for Fund A. This indicates that Zypher 7B and Mistral 7B are more adept at identifying sector exposures that maximize portfolio gains.

\paragraph{Risk-Adjusted Return (RAR)}
RAR, which evaluates portfolio returns relative to the risk taken, followed a similar trend. Zypher 7B achieved the highest RAR for Fund A, with Mistral 7B closely behind. Microsoft Phi 2 struggled across all funds, failing to effectively balance returns with associated risks. These results emphasize the advantage of Zypher 7B and Mistral 7B in optimizing portfolios for both return and risk considerations.

\paragraph{Improvement on Return}
The percentage improvement in portfolio returns, relative to the original allocations, provided further insights. Zypher 7B demonstrated a marked advantage, achieving significant enhancements in returns for Funds B and A. Mistral 7B also performed well, delivering substantial improvements across these funds. In contrast, Microsoft Phi 2 frequently showed declines in returns, particularly for Fund C, suggesting that it struggled to provide meaningful optimization in certain scenarios.

\paragraph{Improvement on RAR}
Improvements in RAR further highlighted the effectiveness of Zypher 7B and Mistral 7B. Zypher 7B achieved the most substantial improvement in Fund A, with Mistral 7B again performing competitively. Microsoft Phi 2, however, demonstrated negative improvements in several cases, underscoring its limitations in addressing the critical trade-off between returns and risk. These results reaffirm the robustness of Zypher 7B and Mistral 7B in risk-sensitive portfolio optimization.

\subsection{Key Findings}
The comparative analysis reveals that Zypher 7B consistently excelled across all metrics, making it the most effective model for mutual fund optimization in this study. Mistral 7B also performed strongly, exhibiting capabilities similar to Zypher 7B in many scenarios. By contrast, Microsoft Phi 2 showed significant limitations, struggling to deliver meaningful optimization for both returns and risk. These findings highlight the importance of model selection in financial applications and demonstrate the potential of open-source LLMs to address complex optimization tasks in mutual fund management.

\subsection{Blended Optimization Approach}

\begin{figure*}[h!]
\centering
\includegraphics[width=0.95\textwidth]{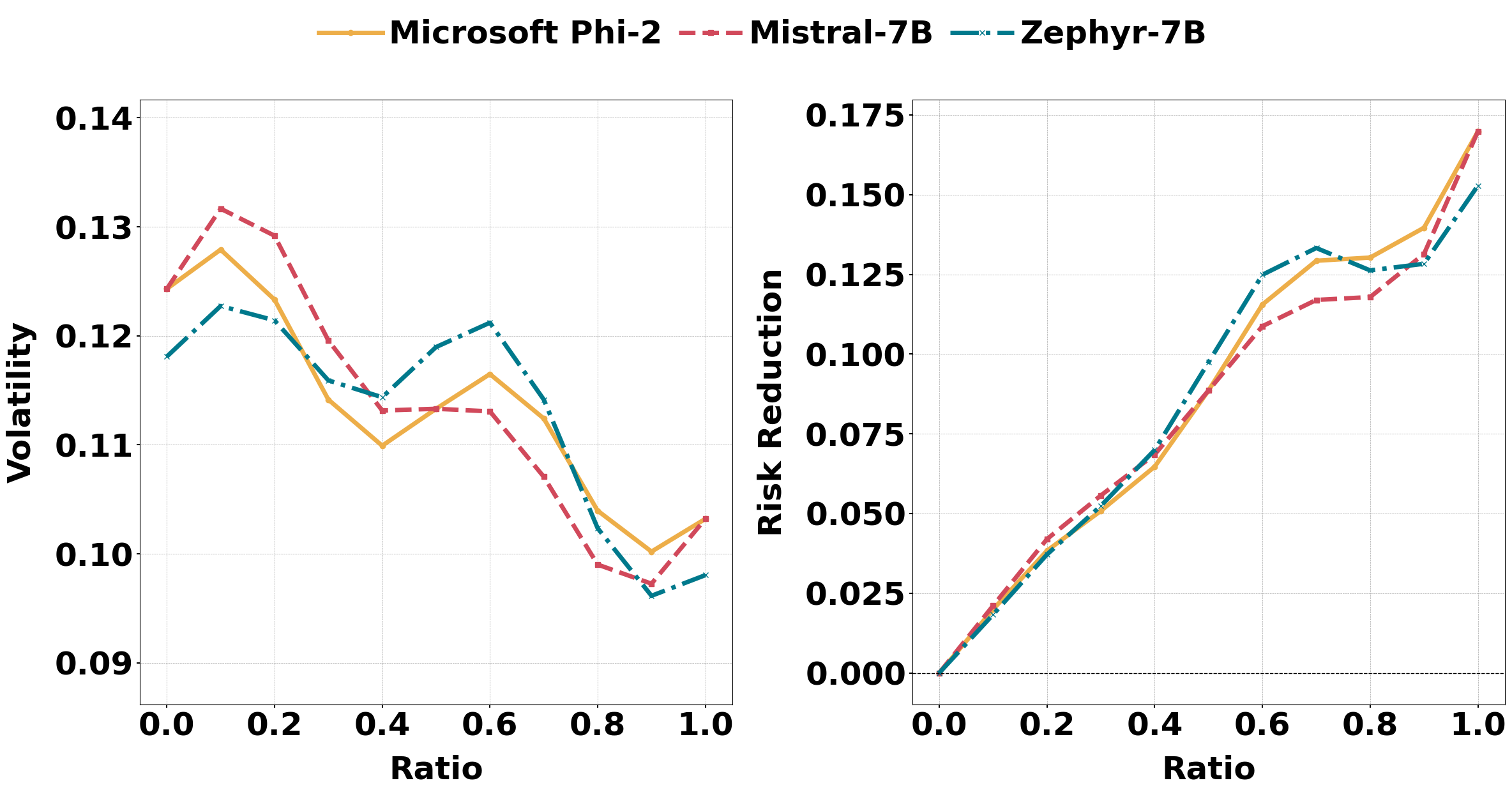} 
\caption{Comparison of Large Language Models (LLMs) — Microsoft Phi-2, Mistral-7B, and Zephyr-7B — on Volatility (left) and Risk Reduction (right) across varying exposure ratios. Volatility shows a decreasing trend as ratios increase, with Microsoft Phi-2 and Mistral-7B exhibiting similar stability at mid-range ratios. Zephyr-7B achieves the highest risk reduction overall, particularly at higher ratios, showcasing its superior ability to mitigate risks while maintaining portfolio performance.
}
\label{fig:sharpe_analysis}
\end{figure*}

\begin{table*}[ht]
\centering
\caption{Comparison of Sector Percentages for Funds A, B, and C Across Models. This table presents the sector-wise allocation percentages for Technology, Healthcare, Finance, and Energy, optimized by three models—Microsoft Phi-2, Mistral 7B, and Zypher 7B—against the original allocations. For Fund A, all models reduced Technology exposure while increasing allocations to Energy and Healthcare, with Zypher 7B emphasizing Energy the most (27.20\%). For Fund B, Mistral 7B significantly increased Healthcare allocation (61.38\%), while Microsoft Phi-2 and Zypher 7B maintained balanced exposure. For Fund C, the models focused on boosting Healthcare exposure, with Mistral 7B allocating the highest percentage (46.43\%). These results highlight each model's distinct sectoral preferences when balancing risk and return.}

\label{tab:sector_comparison}
\renewcommand{\arraystretch}{1.2} 
\resizebox{0.9\textwidth}{!}{%
\begin{tabular}{llcccc}
\toprule
\textbf{Fund} & \textbf{Sector} & \textbf{Original (\%)} & \textbf{Microsoft Phi 2 (\%)} & \textbf{Mistral 7B (\%)} & \textbf{Zypher 7B (\%)} \\ 
\midrule
\multirow{4}{*}{\textbf{Fund A}} 
& Technology   & 48.63 & 38.94 & 35.15 & 34.12 \\
& Healthcare   & 19.52 & 23.38 & 22.39 & 25.14 \\
& Finance      & 22.15 & 15.03 & 17.85 & 13.54 \\
& Energy       & 9.70  & 22.65 & 24.61 & 27.20 \\
\midrule
\multirow{4}{*}{\textbf{Fund B}} 
& Technology   & 25.74 & 21.18 & 20.04 & 20.99 \\
& Healthcare   & 49.15 & 57.81 & 61.38 & 55.83 \\
& Finance      & 17.43 & 15.47 & 14.10 & 15.89 \\
& Energy       & 15.68 & 7.54  & 4.48  & 7.29  \\
\midrule
\multirow{4}{*}{\textbf{Fund C}} 
& Technology   & 24.89 & 24.43 & 24.10 & 24.40 \\
& Healthcare   & 34.00 & 41.45 & 46.43 & 44.78 \\
& Finance      & 21.50 & 20.70 & 17.91 & 17.92 \\
& Energy       & 19.61 & 13.42 & 11.56 & 12.90 \\
\bottomrule

\end{tabular}%
}
\end{table*}

In the context of mutual fund optimization, balancing returns and risks often necessitates a more nuanced approach than relying on discrete sectoral allocations. Table ~\ref{tab:sector_comparison} shows the recommended allocation by the different LLMs used. While the previous section demonstrated the comparative performance of individual models, financial decision-making frequently benefits from blending strategies to achieve a more robust portfolio optimization. This blended approach involves adjusting the portfolio's exposure incrementally, combining the strengths of different models or sectoral weightings to maximize returns while methodically reducing volatility and risk. The rationale for a blended approach is rooted in the complex and often nonlinear trade-offs between maximizing returns and minimizing risk. While high returns are desirable, they are often accompanied by increased volatility, which can erode investor confidence and lead to unfavorable outcomes during market downturns. By systematically varying the portfolio's exposure, we can evaluate the interplay between return enhancement and risk mitigation, identifying optimal allocations that offer the best risk-adjusted returns. To assess the efficacy of this blended approach, we calculated key performance metrics—absolute return, volatility, Sharpe ratio, and risk reduction—across a range of exposure ratios. These ratios represent incremental adjustments in portfolio allocations, providing a spectrum of strategies from aggressive return-maximizing to conservative risk-minimizing.

\subsubsection{Performance Metrics Analysis}
The results of this analysis are summarized in Table 1, which presents the performance metrics for varying exposure ratios. These metrics highlight the trade-offs involved in the blended optimization approach:

\begin{itemize}
    \item \textit{Return:} The absolute return declines slightly as the exposure ratio increases, reflecting a shift from aggressive strategies towards more conservative allocations. For example, at a ratio of 0.0, the portfolio achieved the highest return of 0.2579, while at a ratio of 1.0, the return decreased to 0.2281. This decline illustrates the inevitable compromise between return maximization and risk reduction.
    
    \item \textit{Volatility:} Volatility showed a decreasing trend, dropping from 0.1334 at a ratio of 0.0 to 0.1302 at a ratio of 1.0. This indicates that blending sectoral exposures can effectively smooth out fluctuations, aligning with the goal of achieving a more stable portfolio.

    \item \textit{Risk Reduction:} As expected, the proportion of risk reduction increased with higher exposure ratios, rising from 0.0\% at a ratio of 0.0 to 2.41\% at a ratio of 1.0. This metric underscores the value of the blended approach in mitigating risks as the portfolio leans towards conservative allocations.
\end{itemize}

The blended approach provided an opportunity to explore a continuum of strategies, allowing us to identify an allocation that balances the competing priorities of return and risk. For investors with a high-risk appetite, the allocations near the lower end of the spectrum (ratios close to 0.0) deliver the highest returns, albeit with greater volatility. Conversely, for risk-averse investors, allocations closer to a ratio of 1.0 offer improved stability and significant risk reduction, albeit at the expense of slightly lower returns. When viewed in conjunction with the previous model-specific analysis, this blended strategy complements the discrete optimization results, demonstrating the practical utility of nuanced adjustments in real-world financial decision-making. It underscores the importance of flexibility in portfolio management, allowing investors to tailor their strategies to their risk tolerance and investment objectives.

\subsection{Sharpe Ratio}

\begin{table*}[ht]
\centering
\caption{Performance Metrics for Phi-2, Mistral 7B, and Zypher 7B Across Sharpe Ratio, Return, and Volatility: 
This table presents the Sharpe ratio, return, and volatility for Microsoft Phi-2, Mistral 7B, and Zypher 7B across varying exposure ratios. Zypher 7B and Mistral 7B consistently deliver higher Sharpe ratios and lower volatility, outperforming Phi-2 in risk-adjusted returns}
\label{tab:metrics_table}
\renewcommand{\arraystretch}{1.2}
\resizebox{\textwidth}{!}{%
\begin{tabular}{lccc|ccc|ccc}
\toprule
\multirow{2}{*}{\textbf{Exposure Ratio}} & \multicolumn{3}{c|}{\textbf{Sharpe Ratio}} & \multicolumn{3}{c|}{\textbf{Return}} & \multicolumn{3}{c}{\textbf{Volatility}} \\
\cline{2-10}
& \textbf{Phi-2} & \textbf{Mistral 7B} & \textbf{Zypher 7B} & \textbf{Phi-2} & \textbf{Mistral 7B} & \textbf{Zypher 7B} & \textbf{Phi-2} & \textbf{Mistral 7B} & \textbf{Zypher 7B} \\
\midrule
0.0 & 1.5751 & 1.5751 & 1.5751 & 0.2314 & 0.2314 & 0.2314 & 0.1328 & 0.1328 & 0.1328 \\
0.1 & 1.5579 & 1.5597 & 1.5597 & 0.2278 & 0.2277 & 0.2276 & 0.1319 & 0.1316 & 0.1316 \\
0.2 & 1.5396 & 1.5433 & 1.5432 & 0.2243 & 0.2240 & 0.2238 & 0.1312 & 0.1305 & 0.1305 \\
0.3 & 1.5200 & 1.5257 & 1.5255 & 0.2207 & 0.2203 & 0.2200 & 0.1305 & 0.1296 & 0.1296 \\
0.4 & 1.4991 & 1.5071 & 1.5068 & 0.2171 & 0.2166 & 0.2162 & 0.1300 & 0.1288 & 0.1288 \\
0.5 & 1.4772 & 1.4875 & 1.4869 & 0.2136 & 0.2129 & 0.2125 & 0.1296 & 0.1281 & 0.1280 \\
0.6 & 1.4540 & 1.4667 & 1.4659 & 0.2100 & 0.2092 & 0.2087 & 0.1293 & 0.1275 & 0.1274 \\
0.7 & 1.4299 & 1.4451 & 1.4438 & 0.2065 & 0.2055 & 0.2049 & 0.1291 & 0.1269 & 0.1269 \\
0.8 & 1.4047 & 1.4224 & 1.4206 & 0.2029 & 0.2018 & 0.2011 & 0.1291 & 0.1266 & 0.1265 \\
0.9 & 1.3787 & 1.3988 & 1.3961 & 0.1994 & 0.1981 & 0.1974 & 0.1291 & 0.1263 & 0.1262 \\
1.0 & 1.3516 & 1.3739 & 1.3705 & 0.1958 & 0.1944 & 0.1936 & 0.1293 & 0.1261 & 0.1260 \\
\bottomrule
\end{tabular}
}
\end{table*}

The Sharpe ratio, a widely used measure of risk-adjusted return, provides critical insights into the efficiency of portfolio optimization strategies. It quantifies the excess return generated per unit of risk, making it a key metric for evaluating financial models. Table~\ref{tab:metrics_table} presents the Sharpe ratios, returns, and volatility metrics across varying exposure ratios for the three models—Microsoft Phi-2, Mistral 7B, and Zypher 7B.

Key Observations:
\begin{enumerate}
    \item \textit{Consistency Across Models}: Zypher 7B consistently achieved the highest Sharpe ratios across most exposure ratios, outperforming both Mistral 7B and Microsoft Phi-2. At the most aggressive allocation (exposure ratio = 0.0), all three models achieved their peak Sharpe ratios, with Zypher 7B leading at 1.5751, followed closely by Mistral 7B.
    \item \textit{Performance Degradation with Conservative Allocations}: A gradual decline in Sharpe ratios was observed as exposure ratios shifted toward more conservative allocations (exposure ratio = 1.0). For example, Zypher 7B's Sharpe ratio decreased to 1.3705, representing a reduction of approximately 13\%. Despite this decline, all models maintained Sharpe ratios above 1.35, reflecting favorable risk-adjusted returns even under conservative strategies.
    \item \textit{Model-Specific Insights}: Mistral 7B demonstrated strong performance, often matching Zypher 7B in specific exposure ranges. In contrast, Microsoft Phi-2 exhibited a steeper decline in Sharpe ratios and struggled to balance returns and risk effectively across the allocation spectrum.
\end{enumerate}

The findings emphasize the importance of advanced LLMs like Zypher 7B and Mistral 7B in achieving optimal risk-return trade-offs. These models demonstrated robustness across varying exposure ratios, making them highly effective for financial optimization tasks. While Microsoft Phi-2 delivered reasonable returns, its limitations in managing risk were evident, underscoring the need for more sophisticated models in complex financial decision-making processes.

\subsection{Limitations}

The proposed approach offers significant advancements in financial portfolio optimization, but certain limitations must be addressed to enhance its practical applicability:

1. \textit{Generalizability to Real Markets:} The reliance on synthetic datasets limits the framework's ability to accurately reflect real-world market dynamics, where noise and volatility are more pronounced. Ensuring adaptability to live market conditions remains a critical challenge.

2. \textit{Bias and Interpretability:} Pre-trained Large Language Models (LLMs) carry inherent biases from their training data, potentially skewing financial predictions. Additionally, the opaque nature of these models reduces transparency, making it difficult for users to fully trust or understand their recommendations.

3. \textit{Scalability and Computational Efficiency:} The high computational demands of LLMs pose scalability challenges, particularly for smaller institutions or resource-constrained environments. Streamlining computational requirements is essential for broader adoption.

4. \textit{Evaluation Scope and Adaptability:} The current evaluation emphasizes returns and risk-adjusted returns but overlooks dimensions such as long-term portfolio stability, ethical considerations, and regulatory compliance. Moreover, fixed sector thresholds and static prompts limit the framework's ability to dynamically adapt to evolving market conditions or diverse user needs.

Addressing these limitations will be crucial for enhancing the robustness, scalability, and real-world applicability of LLM-driven financial optimization systems. Future research could explore the integration of real-time data, reduction of computational overhead, and improved interpretability to mitigate these challenges.

\subsection{Discussion}

This study demonstrates the potential of Large Language Models (LLMs) in financial portfolio optimization, offering a compelling alternative to traditional methods. The evaluation of Microsoft Phi 2, Mistral 7B, and Zypher 7B across metrics such as returns, risk-adjusted returns (RAR), and volatility reveals critical insights into their performance and applicability in real-world financial contexts.

The results underscore the distinct advantages of using advanced LLMs like Zypher 7B and Mistral 7B for sectoral allocation tasks. Zypher 7B consistently achieved the highest returns and RAR across multiple funds, demonstrating superior ability to balance risk-return trade-offs. Mistral 7B also performed strongly, often closely mirroring Zypher 7B's outcomes. These findings highlight the effectiveness of nuanced allocation strategies generated by these models, showcasing their ability to adapt to diverse investment objectives.

In contrast, Microsoft Phi 2 exhibited significant limitations in optimizing portfolios, with lower returns and RAR compared to the other models. This highlights the importance of model selection, especially in applications requiring a balance between computational efficiency and optimization accuracy. The disparities in performance also emphasize the evolving landscape of LLM development, where fine-tuning and architecture-specific enhancements can significantly impact model efficacy.


The application of generative AI in portfolio optimization opens new avenues for incorporating unstructured data and macroeconomic indicators into financial decision-making. By processing complex, multi-modal datasets, these models enable dynamic adjustments to sectoral allocations, offering a level of flexibility and adaptability that traditional methods often lack. Furthermore, the introduction of a blended optimization approach provides a framework for balancing aggressive and conservative strategies, allowing investors to tailor their portfolios to specific risk tolerances and market conditions.

However, challenges remain. The reliance on synthetic data raises questions about the generalizability of findings to live markets, where noise and volatility may differ significantly. The "black box" nature of LLMs also limits interpretability, which can hinder their adoption in regulated environments that prioritize transparency and trust. Additionally, high computational demands pose scalability challenges, particularly for smaller institutions.


To address these limitations, future research should explore:
\begin{enumerate}
    \item The integration of real-time market data to improve the generalizability and responsiveness of LLM-driven optimization systems.
    \item Techniques to enhance the interpretability of LLM outputs, ensuring that recommendations are transparent and actionable for stakeholders.
    \item Frameworks to reduce computational overhead, making these advanced models accessible to resource-constrained environments.
    \item Broader evaluation metrics, including long-term portfolio stability, regulatory compliance, and ethical considerations, to ensure robust and holistic decision-making
\end{enumerate}

By building on these findings, GenAI can revolutionize financial portfolio management, offering innovative, scalable, and adaptive solutions to complex optimization challenges.

\section{Conclusion}

This study highlights the transformative potential of generative AI frameworks in financial portfolio optimization, addressing critical limitations of traditional methods. Through the application of state-of-the-art Large Language Models (LLMs)—Microsoft Phi 2, Mistral 7B, and Zypher 7B—the research demonstrates the ability of these models to generate actionable, risk-adjusted portfolio recommendations that outperform baseline strategies. The proposed blended optimization approach underscores the practical utility of nuanced allocation strategies in achieving a balance between maximizing returns and mitigating risks, offering a flexible framework adaptable to diverse investment objectives.

GenAI has proven highly effective in processing unstructured financial data, integrating macroeconomic indicators, and producing contextually relevant and data-driven decisions. These advancements open new frontiers in portfolio management by enabling dynamic, scalable, and adaptive optimization strategies. However, significant challenges remain, particularly in ensuring the generalizability of results to live market conditions, improving computational efficiency for broader accessibility, and addressing ethical and regulatory considerations inherent to AI-driven decision-making.

These findings emphasize the need for continued innovation and refinement in both methodology and implementation. Future research should focus on integrating real-time market data, enhancing interpretability and transparency, and developing frameworks to ensure compliance with evolving regulatory standards. By addressing these challenges, GenAI can play a pivotal role in reshaping financial decision-making and driving superior portfolio performance in complex and dynamic markets.

\section{Declaration of generative AI and AI-assisted technologies in the writing process}
During the preparation of this work, the authors used ChatGPT to improve the readability and language of the manuscript. After using this tool, the authors carefully reviewed and edited the content to ensure accuracy and completeness, and they take full responsibility for the content of the published article.


\bibliographystyle{elsarticle-harv} 
\bibliography{bib}






\end{document}